\documentclass[12pt,a4paper]{article}
\usepackage[dvips]{graphicx}
\usepackage{amsmath,amsthm,amsfonts,amssymb,cite}
\usepackage[notcite,notref]{}
\usepackage[T2A]{fontenc}
\usepackage{float}
\usepackage[cp1251]{inputenc}
\usepackage[russian,english]{babel}
\usepackage{etoolbox}
\usepackage{hyperref}

\DeclareSymbolFont{bletters}{OML}{cmm}{bx}{it}
\DeclareMathSymbol{\bla}{\mathord}{bletters}{'025}
\DeclareMathSymbol{\bmu}{\mathord}{bletters}{'026}

\def \la{\lambda}

\begin{document}

\title{\vspace{-4cm}
	$${}$$\\
	{\bf Dynamic correlation functions of the generalized Tavis-Cummings model}}
\author{{\bf N.~M.~Bogoliubov$^{1,2}$,
		I.~Ermakov$^{2,3}$, A.~Rybin$^2$}\\[0.5cm]
	{\it\small 1. St.-Petersburg Department of V.~A.~Steklov Mathematical Institute RAS}\\
	{\it\small Fontanka 27, St.-Petersburg,
		191023, Russia}\\
	{\it\small 2. ITMO University, Kronverkskiy  49, 197101, St.Petersburg, Russia.}\\
	{\it\small 3. Saint Petersburg State University, University Embankment, 7-9, St Petersburg, Russia.} }

\date{}

\maketitle
\hrule
{\small \begin{abstract} \noindent
The model describing interaction between two-level atoms and a single mode field
in an optical cavity enclosed by a medium with Kerr nonlinearity is considered in our paper. We study the model within the framework of the analytical solution obtained by application of the quantum inverse method (QIM).
The dynamic correlation functions are calculated and  transition elements of photons are represented in the determinant form. The obtained answers depend on the solutions of Bethe equations. We provide the numerical solutions of these equations for the different parameters of the model and study the behaviour of certain dynamical correlation functions.
\end{abstract}}
\vskip0.5cm
\hrule

\vskip1.5cm

\section{Introduction}
The exactly solvable models of quantum nonlinear optics \cite{braak, batch, babel, rybin, wvrk,Kulish1} allow to study the behaviour of
strongly correlated systems in ways that otherwise would
be impossible.



One of the most fundamental exactly solvable for its eigenstates and eigenvalues  models in cavity quantum electrodynamics is the  Jaynes-Cummings (JC) model \cite{jc}.
It  describes the interaction of the single mode of a cavity field with the two-level atom. In spite of its simplicity the model exhibits many nonclassical features caused by its intrinsic nonlinearity, such as, for example, collapses and revivals of the atomic inversion \cite{fractional1,fractional3,fractional4,fractional5}. 

JC model has been thoroughly studied both theoretically \cite{scullyBook} and experimentally \cite{haroch}. The fiftieth anniversary of this model was recently noted \cite{anniversary}. Due to the progress of circuit quantum electrodynamics\cite{circQED} JC model has again attracted physicists's attention\cite{JCrecent}. Many interesting applications of JC model concerned with its generalizations go far beyond its original concept. Natural generalization of the JC model is to consider an ensemble of $N$ noninteracting two-level atoms coupled to a single mode of a cavity field. The model describing such ensemble is known as the Tavis-Cummings (TC) model. It was solved by Tavis and Cummings \cite{tc} at exact resonance, and by Hepp and Lieb for finite detuning \cite{lieb}. TC model may also be expanded in many other ways. For example, one can consider an array of coupled cavities, so-called Jaynes-Cummings-Hubbard model \cite{jch}, or abandon the rotating wave approximation \cite{rwarefuse}.

Lately the models that include either a nonideal cavity or a dynamical Stark shift have again attracted attention \cite{hsk, btb, rs, rm}. Nonlinear effects in cavities can provide new tools in quantum state engeneering\cite{kerr1,kerr2,kerr3}. If the cavity is not ideal an effective
Hamiltonian can be derived \cite{wr, gj, jp}, which adds a fourth-order term in the boson operators to
the simple TC Hamiltonian. We the consider integrable case of such model -- Integrable Generalized Tavis-Cummings (IGTC) model.

The presence of Kerr-like medium changes the behaviour of the system significantly. For example, it crucially changes the Rabi oscillations in JC model\cite{gj,jp}. Kerr-type systems can also be applied to the description of the nonlinear oscillators \cite{nonlinRes1,nonlinRes2} and Bose-Hubbard dimers \cite{milburn,boseHub,nmb}.

The TC model belongs to a class of integrable models known as Gaudin-Richardson systems \cite{gaud, bkul}.  The IGTC model belongs to a different set of integrable models connected with the so called $XXX$
rational $R$-matrix.
It was solved exactly in \cite{bbt} by the Quantum Inverse Method (QIM) \cite{fad,kbi,Kulish2}. The QIM allows us to obtain the exact expressions for the energy spectrum of the model and it's dynamical correlation functions. In this work we show that by using the so-called determinant representation \cite{korepinnorm} it becomes possible to obtain some manageable expressions for correlation functions.

In the QIM approach it is considered that the model is solved if the analytical expression for the eigenenergies, eigenvectors and correlation functions are expressed through the solutions of Bethe equations. The Bethe equations being a set of coupled nonlinear algebraic equations depend on the model under consideration. Because of it's nonlinearity Bethe equations are not easy to solve neither analytically nor even numerically. Nowadays several approaches to this problem have been developed, for example \cite{odeim}, which were successfully applied to the investigation of the Bethe equations of the Goden and Richardson models \cite{gaudinmodel}. For XXX Heisenberg chain with spin-$\frac{1}{2}$ another technique was developed -- the theory of deformed strings (for its extensive review see \cite{strings}).
 In this paper we present some exact numerical solutions of the Bethe equations for the IGTC model in the TC limit. 


The paper is organized in the following way. In the second section we provide the description of the IGTC model as a generalization of the TC model. In the third section we discuss in details the application of the QIM to the model. We present the Bethe equations, the eigenvectors and find the spectrum of the model. In the forth section the analytical expression for the dynamical correlation functions are obtained and presented in determinant form. In the fifth one the numerical analysis of the Bethe equations and correlation functions is presented in details. In the Appendix we provide a number of explicit examples of the solutions of the Bethe equations and present the numerical values of all the quantities we need in the calculation.

\section{Integrable Generaized Tavis-Cummings model}

As the starting point we consider an ensemble of $N$ two-level non interacting atoms of one sort in the cavity. We assume that only one mode of the cavity field has to be taken into account, whereas the other ones are suppressed. Each atom has ground $|\psi^-_i\rangle$ and exited $|\psi^+_i\rangle$ states. 
The TC model model is defined by the Hamiltonian (in the units $\hbar=c=1$):

\begin{equation}\label{tc}
\mathbf{H}_{TC}=\omega a^\dagger a+\omega_0 S^z+g(a^\dagger S^-+aS^+),
\end{equation}
where $\omega$ and $\omega_0$ are frequencies of the cavity mode and the atomic system, $g$ is the cavity-atom coupling constant. $a$ and $a^\dagger$ are the usual annihilation and creation operators for the cavity field, which satisfy usual commutation relations $[a,a^\dagger]=1$. We also introduced the collective $N$-atom Dicke operators $S^\pm,S^z$ (spin operators for which total spin $S\leqslant N/2$) as:

\begin{equation}\label{spinOps}
S^\pm\equiv\sum^N_{i=1}S^\pm_i=\sum^N_{i=1}\sigma^\pm_i\;, \qquad \qquad S^z\equiv\sum^N_{i=1}S^z_i=\frac{1}{2}\sum^N_{i=1}\sigma^i_z\;,
\end{equation}
where
\begin{equation}\label{dikke1}
S^+_i=|\psi^+_i\rangle\langle\psi^-_i|\;, \qquad S^-_i=|\psi^-_i\rangle\langle \psi^+_i|\;, \qquad S^z_i=\frac{1}{2}(|\psi^-_i\rangle\langle \psi^-_i|-|\psi^+_i\rangle\langle \psi^+_i|)\;,\\
\end{equation}
where $\sigma^\pm_i=\frac{1}{2}(\sigma^x\pm i\sigma^y),$ and $\sigma^x_i,\sigma^y_i,\sigma^z_i$ are the Pauli matrices acting each on $i$-th site. Operators $S^z$ and $S^\pm$ satisfy the commutation relations:

\begin{equation}\label{su2}
[S^+,S^-]=2S^z,\,\,[S^z,S^{\pm}]=\pm S^{\pm},
\end{equation}
of the $su(2)$ algebra.

In the Hamiltonian (\ref{tc}) the first term describes single quantized mode of the cavity field, the second one describes the atomic inversion of the whole system, and the first term of the interaction part describes an atomic transition from the excited state to the ground state accompanied by the emission of a photon, whereas the second one describes the reverse process.
The number of excitations $\mathbf{M}$ and the Casimir operator $\mathbf{S}^2$:

\begin{equation}\label{no}
{\bf M}\equiv S^z+a^\dagger a,
\end{equation}

\begin{equation}\label{sconst}
{\bf S}^2\equiv S^+S^-+S^z(S^z-1)
\end{equation}
are two nontrivial constants of motion $[\mathbf{H}_{TC},\mathbf{M}]=[\mathbf{H}_{TC},\mathbf{S}^2]=0$.

In this paper we shall consider the exactly solvable extension of the TC model that describes the system in complex environment and takes into account the spin-spin interaction. The Hamiltonian for the IGTC model can be written in the form \cite{bbt}:
\begin{equation}\label{hk}
{\bf H}_K=\omega  a^{\dag}a+\omega_0S^z+g( a^{\dag}S^-+aS^+)+\gamma( a^{\dag} a^{\dag}aa+(S^z)^2).
\end{equation}
Here fourth-order term in the boson operators describes a Kerr-like medium, whereas $(S^z)^2$ describes the 
spin-spin coupling. When $\gamma=0, \; \mathbf{H}_K$ reduces to the TC model (\ref{tc}) and this becomes the JC model \cite{jc} for $N=1 \; (S=\frac{1}{2})$. When $S=\frac{1}{2}, \; (S^z)^2=1$ and $\mathbf{H}_K$ with $\gamma\neq 0$ is the JC model with Kerr nonlinearity \cite{nonlinJC}.

It's easy to verify that operator $\mathbf{M}$ commutes with the Hamiltonian $\mathbf{H}_K : [\mathbf{H}_K,\mathbf{M}]=0$, so we can consider another Hamiltonian $\mathbf{H}=g^{-1}(\mathbf{H}_K+(\gamma-\omega)\mathbf{M}-\gamma\mathbf{M}^2)$ and $[\mathbf{H}_K,\mathbf{H}]=0$. We can then write
\begin{equation}\label{ham}
{\bf H}=\Delta S^z+(a^\dagger S^-+aS^+)+ca^\dagger aS^z,
\end{equation}
where $c=-2g^{-1}\gamma$ is a new effective coupling constant and $\Delta=g^{-1}(\gamma+\omega_0-\omega)$ is the frequency shifted detuning. Note, that the last term on the right-hand side of (\ref{ham}) causes photon number dependent changes in the atomic transitions and describes therefore a Stark shift. Henceforth, we shall consider ${\bf H}$, but the same results can immediately be extended to the model with Kerr nonlinearity, Hamiltonian (\ref{hk}), through the mapping given above.



\section{The determinant representation of correlation functions}

To apply QIM to the solution of the model we consider the two $2\times 2$ matrix operators ${\bf L}_B(\lambda),\, {\bf L}_S(\lambda)$ for bosons and spins respectively \cite{bbt}:

\begin{equation}
{\bf L}_B(\lambda)=\left(
\begin{array}{cc}
\lambda-\Delta -c^{-1}-ca^{\dag}a & a^{\dag}  \\
a & -c^{-1}
\end{array}
\right),  \label{lob}
\end{equation}

\begin{equation}
{\bf L}_S(\lambda)=\left(
\begin{array}{cc}
\lambda - cS^z & cS^+  \\
cS^- & \lambda + cS^z
\end{array}
\right),  \label{los}
\end{equation}
in which $\lambda$ is a complex number. Note that the elements of ${\bf L}_B(\lambda)$ and   ${\bf L}_S(\lambda)$ mutually commute. For the monodromy matrix of the QIM  $T(\lambda)$ we can now set
\begin{equation}  \label{mon}
{\bf T}(\lambda )={\bf L}_B(\lambda ){\bf L}_S(\lambda )=\left(
\begin{array}{cc}
{\bf A}(\lambda ) & {\bf B}(\lambda ) \\
{\bf C}(\lambda ) & {\bf D}(\lambda )
\end{array}
\right) ,
\end{equation}
so that
\begin{eqnarray}
{\bf B}(\lambda)&=& \lambda {\bf X}-{\bf Y}; \label{b} \\
\nonumber {\bf X} &=& a^{\dag}+cS^+,\,{\bf Y}=(1+c\Delta)S^+-ca^{\dag}S^z+c^{2}a^{\dag}aS^+;  \\
\nonumber  [{\bf X},{\bf Y}] &=& 0,
\end{eqnarray}
and
\begin{equation}\label{c}
{\bf C}(\lambda)=\lambda a -(S^-+caS^z)\equiv\lambda a -{\bf Z}.
\end{equation}
We also have\begin{eqnarray}
{\bf A}(\lambda) &=& (\lambda-\Delta -c^{-1}-ca^{\dag}a)(\lambda - cS^z)+ca^\dag S^-\,, \label{ad}\\
\nonumber           {\bf D}(\lambda) &=& caS^+-c^{-1}(\lambda+cS^z).
\end{eqnarray}

Multiplying $L$-operators in another than in (\ref{mon}) order we obtain the following expression for the
monodromy matrix
\begin{equation}  \label{mont}
\widetilde{{\bf T}}(\lambda )={\bf L}_S(\lambda ){\bf L}_B(\lambda )=\left(
\begin{array}{cc}
\widetilde{{\bf A}}(\lambda ) & \widetilde{{\bf B}}(\lambda ) \\
\widetilde{{\bf C}}(\lambda ) & \widetilde{{\bf D}}(\lambda )
\end{array}
\right) .
\end{equation}

The entries of this matrix satisfy the involution relations \cite{nmb}
\begin{equation}\label{coper}
\widetilde{\bf B}(\lambda)={\bf C}^+(\lambda^*),\,\, \widetilde{\bf C}(\lambda)={\bf B}^+(\lambda^*).
\end{equation}

The introduced monodromy matrices satisfy the intertwining relation

\begin{eqnarray}
  \mathbf{R}(\lambda,\mu)\mathbf{T}(\lambda)\mathbf{T}(\mu) = \mathbf{T}(\mu)\mathbf{T}(\lambda)\mathbf{R}(\lambda,\mu) \\
\nonumber  \mathbf{R}(\lambda,\mu)\tilde{\mathbf{T}}(\lambda)\tilde{\mathbf{T}}(\mu) = \tilde{\mathbf{T}}(\mu)\tilde{\mathbf{T}}(\lambda)\mathbf{R}(\lambda,\mu)
\end{eqnarray}


with the rational $R$-matrix
\begin{equation}  \label{R}
{\bf R}(\lambda ,\mu )=\left(
\begin{array}{cccc}
f(\mu ,\lambda ) & 0 & 0 & 0 \\
0 & g(\mu ,\lambda ) & 1 & 0 \\
0 & 1 & g(\mu ,\lambda ) & 0 \\
0 & 0 & 0 & f(\mu ,\lambda )
\end{array}
\right) ,
\end{equation}
with the enries
\[
f(\mu ,\lambda )=1-\frac c{\mu -\lambda },\;\;\;\;\;\;\;\;g(\mu ,\lambda
)=-\frac c{\mu -\lambda }.
\]

The traces of the monodromy matrices (\ref{mon})and (\ref{mont}) coinside:
\[
Tr {\bf T}(\lambda)=Tr  \widetilde{\bf T}(\lambda)=\tau(\lambda)\,,
\] and commute for arbitary complex numbers $\lambda, \mu$:

\begin{equation}\label{trcom}
[\tau (\lambda ),\tau (\mu )]=0.
\end{equation}

In the explicit form:

\begin{equation}  \label{tre}
\tau (\lambda )=\lambda ^2-\lambda (c{\bf M+}\Delta +2c^{-1})+c{\bf H},
\end{equation}
where   ${\bf M}$ is the number operator (\ref{no}) and ${\bf H}$ is the  Hamiltonian (\ref{ham}). It can be checked that
\begin{eqnarray}
\label{ht}  {\bf H}&=& c^{-1}\tau (0), \\
\nonumber {\bf M} &=& - c^{-1}\frac{\partial \tau (\lambda)}{\partial \lambda} \mid_{\lambda=0}-c^{-1}\Delta-2c^{-2}.
\end{eqnarray}


It can be shown that $\mathbf{M}\mathbf{B}(\lambda)=\mathbf{B}(\lambda)(\mathbf{M}+1)$, and likewise $\mathbf{S}^2\mathbf{B}(\lambda)=\mathbf{B}(\lambda)\mathbf{S}^2$. So $\mathbf{B}(\lambda)$ acts as a creation operator for the quasi-particles,
while $\mathbf{C}(\lambda)$ is an annihilation operator.

The M-particle state vectors are constructed in the usual fashion for the QIM method
\begin{equation}  \label{wavef}
{\bf \mid \Psi }_{S,M}(\{\bla\})\rangle
=\prod_{j=1}^M{\bf B}(\lambda_j)\mid \Omega _S\rangle=\prod_{j=1}^M(\lambda_j {\bf X}-{\bf Y})\mid \Omega _S\rangle,
\end{equation}
where the vacuum state $|\Omega _S\rangle=|0\rangle |S,-S\rangle$ ($a|0\rangle=0; S^-|S,-S\rangle=0$, with ${\bf S}^2|S,-S\rangle=S(S+1)|S,-S\rangle$, and $S^z|S,-S\rangle=-S|S,-S\rangle$)
is annihilated by operator $\bf {C}$
\[
{\bf C}(\lambda )\mid \Omega _S\rangle =0\,,
\]
and is the eigenstate of the operators ${\bf A,D}$:
\begin{equation}\label{ADvac}
{\bf A}(\lambda )\mid \Omega _S\rangle =a(\lambda )\mid \Omega
_S\rangle\,, \,\,\,\,
{\bf D}(\lambda )\mid \Omega _S\rangle =d(\lambda )\mid \Omega
_S\rangle \,,
\end{equation}
where
\begin{equation}\label{AD}
a(\lambda )=(\lambda -\Delta -c^{-1})(\lambda +cS)\,, \,\,\,\,
d(\lambda )=-c^{-1}(\lambda -cS)\,.
\end{equation}
In formula \ref{wavef} a short-hand notation $\{{\bf x}\}\equiv \{\la_1,\la_2, \ldots,\la_M\}$ is used.

In the explicit form the state vector (\ref{wavef}) has the form
\begin{equation}\label{wavefef}
{\bf \mid \Psi }_{S,M}(\{\bla\})\rangle
=\sum_{m=0}^M (-1)^{M-m}e_m {\bf X}^m{\bf Y}^{M-m}\mid \Omega _S\rangle,
\end{equation}
where
\[
e_m = \sum_{i_1<i_2<\ldots <i_m} \lambda_{i_1} \lambda_{i_2}\ldots \lambda_{i_m}
\]
is the elementary symmetric function \cite{macd}.

The conjugated M-particle state vectors are equal to
\begin{eqnarray}
\langle{\bf \Psi }_{S,M}(\{\bla \})\mid &=& \langle \Omega _S\mid \prod_{j=1}^M{\bf C}(\la_j)=\langle \Omega _S\mid\prod_{j=1}^M [\la_j a -(S^-+caS^z)] \label{cwavef} \\
\nonumber  &=& \langle \Omega _S\mid\sum_{m=0}^M (-1)^{M-m}e_m a^m {\bf Z}^{M-m}
\end{eqnarray}
and
\[
\langle \Omega _S\mid{\bf B}(\la ) =0.
\]
By the construction the state vectors (\ref{wavef}) and (\ref{cwavef}) are symmetric functions of their arguments $\{\bla\}\equiv\{\la_1,\la_2,...,\la_n\}$.

The state vectors (\ref{wavef}\ref{cwavef}) are eigenvectors of the transfer matrix, and thus of the Hamiltonian if $\{\bla \}$ are the roots of the Bethe equations which here take the form, for $n=1,2,\ldots,M$,
\begin{equation}  \label{bethe}
(1+\Delta c-c\lambda _n)\frac{\lambda _n+cS}{\lambda
	_n-cS}=\prod_{ j=1,j\neq n}^M\frac{\lambda _n-\lambda _j+c}{\lambda _n-\lambda _j-c}\,,
\end{equation}
where $0\leq S\leq \frac N2$. Evidently, there are $K=min(2S,M)+1$ (modulo the permutation group) sets of
solutions of these $M$ Bethe equations. The complex valued roots are pairwise conjugated.

The eigenvalues of the transfer matrix (\ref{tre}):
\begin{equation}
 \theta _{S,M}(\mu) =  a(\mu)\prod_{j=1}^M(1-\frac{c}{\mu-\lambda_j})- d(\mu)\prod_{j=1}^M(1+\frac{c}{\mu-\lambda_j}).
\end{equation}

From the equation (\ref{ht}) we see that the $M$-particle eigenenergies of the Hamiltonian (\ref{ham}) are equal to:
\begin{eqnarray}
{\bf H} \mid {\bf \Psi }_{S,M}(\{\bla \})\rangle &=& E_{S,M} \mid {\bf \Psi }_{S,M}(\{\bla \})\rangle, \label{ege} \\
\nonumber   E_{S,M} &=&\frac{S}{c}\prod_{j=1}^M\left(1-\frac{c}{\lambda}_j\right)-\left(S\Delta +\frac{S}{c}\right)\prod_{j=1}^M\left(1+\frac{c}{\lambda}_j\right).
\end{eqnarray}
The ground state of the Hamiltonian (\ref{ham}) corresponds to the minimal value of eigenenergy.
The set of solutions of Bethe equations that defines this state will be labeled by $\sigma_g$: $\{{\bla}^{\sigma_g}\}$.

The form of the Bethe equations drastically depends on the parameter $c$. In case of vanishing nonlinearity $c\rightarrow 0$ the Bethe equations (\ref{bethe}) transform to another set of equations (\ref{bethetc}), which refers to the well-studied TC-model \cite{gaud,bkul}.

\begin{equation}\label{bethetc}
\frac{2S}{\tilde{\lambda}_n}-\tilde{\lambda}_n+\Delta=\sum^M_{j=1,j\neq n}\frac{2}{\tilde{\lambda}_n-\tilde{\lambda}_j}.
\end{equation}

The energy spectrum of the TC-model can be expressed in the following form:

\begin{equation}\label{eigentc}
\mathcal{E}_{S,M}^\sigma=-S\left(\Delta+2\sum^M_{j=1}\frac{1}{\tilde{\lambda}^\sigma_j}\right).
\end{equation}

Consider the state vectors constructed by operators (\ref{coper})
\begin{equation}\label{stvc}
\mid \widetilde{{\bf \Psi }}_{S,M}(\{\bla\})\rangle=\prod_{j=1}^M\widetilde{{\bf B}}(\lambda_j)\mid \Omega _S\rangle ,\,\,
\langle\widetilde{{\bf \Psi }}_{S,M}(\{\bla \})\mid = \langle \Omega _S\mid \prod_{j=1}^M\widetilde{{\bf C}}(\la_j) \,.
\end{equation}
It was proved in \cite{kbi} that on the solutions of Bethe equations
\begin{equation}\label{prop}
\mid \widetilde{{\bf \Psi }}_{S,M}(\{\bla\})\rangle=\nu_M \mid {\bf \Psi }_{S,M}(\{\bla\})\rangle ,\,\,
\langle\widetilde{{\bf \Psi }}_{S,M}(\{\bla \})\mid=\nu^{-1}_M\langle{\bf \Psi }_{S,M}(\{\bla \})\mid\,,
\end{equation}
where for the model under consideration
\begin{equation}\label{nu}
\nu^{-1}_M=\prod_{j=1}^M(1+\Delta c-c\lambda_j)=\prod_{j=1}^M\frac{\lambda_j-cS}{\lambda_j+cS}\,.
\end{equation}

Note, that from Bethe equations (\ref{bethe}) it follows that
\begin{equation}\label{inv}
\prod_{n=1}^M(1+\Delta c-c\lambda _n)\frac{\lambda _n+cS}{\lambda
	_n-cS}=1\,.
\end{equation}

%

In \cite{bbt} it was proved that Bethe state vectors form a complete orthogonal set
\begin{multline}\label{orth}
\langle{\bf \Psi }_{S,M} (\{{\bla}^{\sigma_1} \}) \mid {\bf \Psi }_{S,M}(\{{\bla}^{\sigma_2} \})\rangle\sim\delta_{\sigma_1,\sigma_2} \\
\sum_\sigma\frac{\mid {\bf \Psi }_{S,M}(\{{\bla}^{\sigma} \})\rangle \langle{\bf \Psi }_{S,M} (\{{\bla}^{\sigma} \}) \mid}{\mathcal{N}_\sigma^2}=\mathbb{I}\,.
\end{multline}
Index $\sigma$ denotes the independent sets of solutions of Bethe equations (\ref{bethe}), and the summation is over all $K$ sets of solutions.

The scalar product of the Bethe state vectors for $M=1,2$ can be calculated explicitly:

\begin{equation}\label{norm1full}
S_1(\mu,\lambda) \equiv  \langle{\bf \Psi }_{S,1}(\mu)|{\bf \Psi }_{S,1}(\lambda)\rangle=-c^2s^2+s(2+c(2\Delta-\lambda-\mu))+\lambda\mu,
\end{equation}
and
\begin{multline}
S_2(\mu_1,\mu_2,\lambda_1,\lambda_2) \equiv  \langle{\bf \Psi }_{S,2}(\mu_1,\mu_2)|{\bf \Psi }_{S,2}(\lambda_1,\lambda_2)\rangle=\label{norm2full}\\
2 c^4 s^4-4 c^4 s^3+2 c^4 s^2-\left(\lambda _1+\lambda _2\right) \left(-2 c^3 s^3+2 c^3 s^2+c \left(4 s^2-2 s\right) (c \Delta +1)\right)-\\
\left(\mu _1+\mu _2\right) \left(-2 c^3 s^3+2 c^3 s^2+4 c s^2 (c \Delta +1)-2 c s (c \Delta +1)\right)+4 c^2 s^2 (c \Delta +1)+2 c^2 \lambda _1 \lambda _2 s^2+\\
2 c^2 \mu _1 \mu _2 s^2+(c \Delta +1) \left(-8 c^2 s^3+4 c^2 s^2-2 c^2 s+\left(8 s^2-4 s\right) (c \Delta +1)\right)+\\
2 \left(\lambda _1+\lambda _2\right) \left(\mu _1+\mu _2\right) s (c \Delta +1)-2 c \left(\lambda _1+\lambda _2\right) \mu _1 \mu _2 s-\\
2 c \lambda _1 \lambda _2 \left(\mu _1+\mu _2\right) s+2 \lambda _1 \lambda _2 \mu _1 \mu _2.
\end{multline}
The further straight forward calculation of scalar products for $M>2$ is rather difficult because the number of commutation relations, one needs to evaluate in order to get the answer, grows exponentially.

To obtain the analytical expression for the correlators for the arbitrary $M$ we shall use the Slavnov's formula for the scalar products \cite{sl,kit,kit2}. This formula adopted for the model under consideration has the form:
\begin{eqnarray}
S_M(\{\bmu\},\{\bla\})&=&  \langle{\bf \Psi }_{S,M}(\{\bmu \})\mid {\bf \Psi }_{S,M}(\{\bla \})\rangle \label{scpr} \\
\nonumber &=&\frac{c^M \prod^M_{j=1}d(\lambda_j)}{\prod_{j>k}(\mu_k-\mu_j)\prod_{\alpha<\beta}(\lambda_\beta-\lambda_\alpha)}
\det T(\{\bmu\},\{\bla\}),
\end{eqnarray}
where the entries of the $M\times M$ matrix $T(\{\bmu\},\{\bla\})$ are
\begin{multline} \label{entr}
T_{ab}\equiv T(\mu_a,\la_b) = -c^{-1}\frac{\partial}{\partial \lambda_b}\theta(\mu_a,\{\bla\})  \\
 = \frac{1}{(\lambda_b-\mu_a)^2}\left\lbrace a(\mu_a)\prod^M_{j=1,j\neq b}\left(1-\frac{c}{\mu_a-\lambda_j}\right)-d(\mu_a)\prod^M_{j=1,j\neq b}\left(1+\frac{c}{\mu_a-\lambda_j}\right)\right\rbrace\,.
\end{multline}
It is supposed here and below that $\{\bla\}$ are the solutions of Bethe equations, while $\{\bmu\}$ is the set of arbitrary parameters.
Functions $a(\mu),d(\mu)$ are defined in (\ref{ADvac}).

The square of the norm of the eigenvectors is calculated by the Gaudin formula \cite{gaud, korepinnorm} which is (\ref{scpr}) in the limit $\{\bmu\}\rightarrow\{\bla\}$:
\begin{equation}\label{norm}
\mathcal{N}^{\,2}=S_M(\{\bla\},\{\bla\})=c^M\prod_{j=1}^M d^2(\lambda_j)\prod_{\alpha\neq\beta}\frac{\lambda_\alpha-\lambda_\beta+c}{\lambda_\alpha-\lambda_\beta} \det \Phi(\{\bla\}),
\end{equation}
where the entries of the $M\times M$ matrix $\Phi$ are
\begin{equation}\label{f}
\Phi_{ab}=-\frac{\partial}{\partial\lambda_b} \ln\left\lbrace \frac{a(\lambda_a)}{d(\lambda_a)}\prod_{k=1,k\neq a}^M\frac{\lambda_a-\lambda_k-c}{\lambda_k-\lambda_a-c}\right\rbrace\,.
\end{equation}

The determinant representation (\ref{scpr}) may be used in calculation of the transition element of the photon annihilation operator
\begin{equation} \label{trel}
\langle{\bf \Psi }_{S,M-n} (\{\bla' \})\mid a^n \mid {\bf \Psi }_{S,M}(\{\bla \})\rangle\,,
\end{equation}
where $\{\bla \}$ and $\{\bla' \}$ are the solutions of Bethe equations (\ref{bethe}) for $M$ and $M-n$ particles respectively. Really, notice that from the definition (\ref{c}) it follows that $\lim_{\lambda \rightarrow \infty} \lambda^{-1} C(\lambda)=a$ and hence
\begin{equation}\label{aslim}
\langle{\bf \Psi }_{S,M-n} (\{\bmu \})\mid a^n \mid {\bf \Psi }_{S,M}(\{\bla \})\rangle=\lim_{\mu_1,\mu_2,\ldots,\mu_n \rightarrow \infty}  \frac{ S_M(\{\bmu\},\{\bla\})}{\mu_1\mu_2\ldots\mu_n}.
\end{equation}
Replacing the arbitrary parameters $\{\bmu \}$ by the solutions of Bethe equations $\{\bla' \}$ we obtain the answer for (\ref{trel}).
Denoting
\[
V_{ab}=\lim_{\mu_a \rightarrow \infty} \frac{ \mu_a^{-M+a-1}}{(a-1)!}\frac{\partial ^{a-1}}{\partial \mu_a^{a-1}} T (\mu_a,\la_b)
\]
we obtain the following answer
\begin{multline}\label{aa}
A_{M,n}(\{\bmu \},\{\bla \})\equiv
\langle{\bf \Psi }_{S,M-n} (\{\bmu \})\mid a^n \mid {\bf \Psi }_{S,M}(\{\bla \})\rangle \\
= c^M\prod^M_{j=1}d(\lambda_j)\frac{\prod_{j=1}^{M-n} \prod_{\alpha=1}^M (\mu_j-\lambda_\alpha)}{\prod_{j>k>n}(\mu_k-\mu_j)\prod_{\alpha<\beta}(\lambda_\beta-\lambda_\alpha)}
\det T_{(n)}(\{\bmu\},\{\bla\})\,.
\end{multline}
Here set $\{\bmu\}\equiv\{\mu_1,\mu_2,...,\mu_{M-n}\}$ has the length $M-n$, whereas the set $\{\bla\}\equiv\{\lambda_1,\lambda_2,...,\lambda_M\}$ has the length $M$. The entries of the $M\times M$ matrix $T_{(n)}$ are equal to $V_{ab}$ for $a\leq n$ and are $T_{ab}$ (\ref{entr}) for $a>n$, $1\leq b \leq M$. To obtain the answer for the transition element (\ref{trel}) we have to change parameters $\{\bmu\}$ on the solutions of Bethe equations for $M-n$ particles.

For instance, for $n=1$ the answer for the transition element is
\begin{multline}\label{a2}
\langle{\bf \Psi }_{S,M-1} (\{{\bla}' \})\mid a \mid {\bf \Psi }_{S,M}(\{\bla \})\rangle \\
= c^M\prod^M_{j=1}d(\lambda_j)\frac{\prod_{j=1}^{M-1} \prod_{\alpha=1}^M (\la'_j-\lambda_\alpha)}{\prod_{j>k>1}(\la'_k-\la'_j)\prod_{\alpha<\beta}(\lambda_\beta-\lambda_\alpha)}
\det T_{(1)}(\{{\bla}'\},\{\bla\})\,,
\end{multline}
where the entries of $M\times M$ matrix
\[
T_{(1)}(\{{\bla}'\},\{\bla\})=
\left(
\begin{array}{ccccc}
-1 & -1 & -1 & \ldots & -1 \\
T_{21} & T_{22} & T_{23} & \ldots & T_{2M} \\
\vdots & \vdots & \vdots &\ddots & \vdots \\
T_{M1} & T_{M2} & T_{M3}& \ldots & T_{MM} \\
\end{array}
\right)\,.
\]

To find the transition element of the creation operator one can take the complex conjugation of the transition element of the annihilation operator and obtain:
\begin{multline}\label{croper}
\langle{\bf \Psi }_{S,M-n} (\{{\bla}' \})\mid a^n \mid {\bf \Psi }_{S,M}(\{\bla \})\rangle^*=
\langle \Omega _S\mid \prod_{j=1}^{M-n}{\bf C}(\lambda'_j)a^n\prod_{j=1}^{M}{\bf B}(\lambda_j)\mid \Omega _S\rangle^* \\
=\langle \Omega _S\mid \prod_{j=1}^{M}{\bf B}^+(\lambda_j)(a^\dag)^n\prod_{j=1}^{M-n}{\bf C}^+(\lambda'_j)\mid \Omega _S\rangle\\
= \langle \Omega _S\mid \prod_{j=1}^{M}\widetilde{{\bf C}}(\lambda_j)(a^\dag)^n\prod_{j=1}^{M-n}\widetilde{\bf B}(\lambda'_j)\mid \Omega _S\rangle\\
=\frac{\nu'_{M-n}}{\nu_{M}}\langle{\bf \Psi }_{S,M} (\{\bla \})\mid (a^\dag)^n \mid {\bf \Psi }_{S,M-n}(\{\bla' \})\rangle\,,
\end{multline}
where the definition (\ref{stvc}) and the properties (\ref{coper}),(\ref{prop}) have been used, and ${\bla'}$ are the solutions of Bethe equations (\ref{bethe}) with $M-n$
particles. The obtained formula allows to express the transition element of the creation operator in the determinantal form:
\begin{multline}\label{aad}
A^\dag_{M,n}(\{\bla \},\{\bla' \})\equiv
\langle{\bf \Psi }_{S,M} (\{\bla \})\mid(a^\dag)^n \mid {\bf \Psi }_{S,M-n}(\{\bla' \})\rangle \\
= \frac{\nu_M}{\nu'_{M-n}} A^*_{M,n}(\{\bmu \},\{\bla \})\,,
\end{multline}
where $A^*_{M,n}(\{\bmu \},\{\bla \})$ is the Complex conjugated coefficient (\ref{aa}) on the solutions of Bethe equations.


The obtained representations for the transition elements (\ref{aa}) and  (\ref{aad})
allow us to calculate different $n$-photon time-dependent correlation functions.

We define the $n$-photon Green function $\langle a^n(t) (a^\dag)^n\rangle_M$ as the average  taken over the $M$-particle ground state of the model $\mid {\bf \Psi }_{S,M}(\{{\bla}^{\sigma_g} \})\rangle$:
\begin{multline}\label{tdcf}
\langle a^n(t) (a^\dag)^n\rangle_M
=\frac{1}{\mathcal{N}_{\sigma_g}^2}
\langle{\bf \Psi }_{S,M} (\{{\bla}^{\sigma_g} \})\mid e^{-i{\bf H}t} a^n e^{i{\bf H}t} (a^\dag)^n \mid {\bf \Psi }_{S,M}(\{{\bla}^{\sigma_g} \})\rangle\\
=\sum_\sigma \frac{e^{it(E^\sigma_{S,M}-E^{\sigma_g}_{S,M}})}{\mathcal{N}^2_{\sigma_g}  \mathcal{N}^2_\sigma }
\langle{\bf \Psi }_{S,M} (\{{\bla}^{\sigma_g} \})\mid a^n \mid {\bf \Psi }_{S,M+n}(\{\bla^\sigma \})\rangle\\
\times  \langle {\bf \Psi }_{S,M+n}(\{\bla^\sigma \})\mid (a^\dag)^n \mid {\bf \Psi }_{S,M}(\{{\bla}^{\sigma_g} \})\rangle\\
=\sum_\sigma \frac{e^{it(E^\sigma_{S,M}-E^{\sigma_g}_{S,M}})}{\mathcal{N}^2_{\sigma_g}  \mathcal{N}^2_\sigma }\Bigl| \langle {\bf \Psi }_{S,M} (\{{\bla}^{\sigma_g} \})\mid a^n \mid {\bf \Psi }_{S,M+n}(\{\bla^\sigma \})\rangle \Bigr|^2 \frac{\nu^{\sigma}_{M+n}}{\nu^{\sigma_g}_M}\,,
\end{multline}
where the sum is taken over $K=\min (2S,M+n)+1$ sets of solutions of Bethe equations (\ref{bethe}) for $M+n$ particles ${\bla^\sigma}$ labeled by $\sigma$\,, and the complete orthogonal set (\ref{orth}) of eigenstates of the Hamiltonian (\ref{ham}) was used.
Substituting formulas (\ref{ege}), (\ref{nu}), (\ref{norm}) and (\ref{aa}) into (\ref{tdcf}) we obtain the final answer.

The developed technique allows us to calculate different dynamical correlation functions of the considered model and respectively of the TC model in the $c\rightarrow 0$ limit.

\section{Time evolution of the atomic inversion}

Knowing the projection of the initial state on the state vectors of the model one can obtain the answers for the correspondent dynamical correlation functions.
Let us consider the time evolution of the coherent states. Let us assume that in the initial state $ \mid \Phi_0\rangle$ the spin system is in the ground state $\mid S,-S\rangle$ ($S^-\mid S,-S\rangle=0$) and the field is in the coherent state $\mid \alpha \rangle $ $\alpha\in\mathbb{C}$ ($ a\mid \alpha \rangle=\alpha\mid \alpha \rangle $)
\begin{equation}\label{initst}
\mid {\bf\Phi}_0\rangle=\mid \alpha \rangle \mid S,-S\rangle=e^{-\frac{|\alpha|^2}{2}}e^{\alpha a^\dag}\mid 0\rangle \mid S,-S\rangle\,.
\end{equation}
The time evolution of the initial state can be obtained by application of the evolution operator $U(t)=exp(i{\bf H} t)$, so  we have $\mid {\bf\Phi}(t)\rangle=U(t)\mid {\bf\Phi}_0\rangle$. We may use representations (\ref{cwavef}) and (\ref{wavefef}) to find the projections of the initial states on the Bethe state vectors:
\begin{multline}\label{cohpro}
\langle{\bf \Psi }_{S,M} (\{\bla \})\mid {\bf\Phi}_0\rangle=(-1)^{M+1}(c\alpha S)^M e^{-\frac{|\alpha|^2}{2}}\sum_{m=0}^Me_m (cS)^{-m}\equiv f_M(\{\bla \})\,, \\
\langle {\bf\Phi}_0\mid {\bf \Psi }_{S,M} (\{\bla \})\rangle=(c\bar{\alpha} S)^M e^{-\frac{|\alpha|^2}{2}}\sum_{m=0}^M (-1)^{M-m} e_m (cS)^{-m}\equiv g_M(\{\bla \})\,.
\end{multline}

The evolution of field intensity can be constructed  through the complete set of eigenfunctions $\mid{\bf \Psi }_{S,M} (\{{\bla}^\sigma \})\rangle$:
\begin{multline}\label{intens}
\langle \Phi(t)\mid (a^\dag)^n a^n \mid \Phi(t)\rangle
=\sum_M\sum_{\sigma_1,\sigma_2,\sigma_3}\frac{e^{it(E_{S,M}^{\sigma_1}-E_{S,M}^{\sigma_3})}}
{\mathcal{N}^2_{\sigma_1}\mathcal{N}^2_{\sigma_2}\mathcal{N}^2_{\sigma_3}}\\
\times f_M(\{{\bla}^{\sigma_1} \})g_M(\{{\bla}^{\sigma_3} \})A_{M,n}^\dag(\{{\bla}^{\sigma_3} \},\{{\bla}^{\sigma_2} \})
A_{M,n}(\{{\bla}^{\sigma_2} \},\{{\bla}^{\sigma_1} \})\,.
\end{multline}
The summation here is over all sets of solutions of correspondent Bethe equations.

Knowing the answer for the field intensity we can find the evolution of atomic inversion applying equality
\begin{multline}\label{atinv}
\langle \Phi(t)\mid S^z \mid \Phi(t)\rangle=\langle \Phi(t)\mid {\bf M}\mid \Phi(t)\rangle -\langle \Phi(t)\mid a^\dag a \mid \Phi(t)\rangle  \\
=<n>-S-\langle \Phi(t)\mid a^\dag a \mid \Phi(t)\rangle \,,
\end{multline}
where $\bf M $ is the number operator (\ref{no}), and $<n>=|\alpha|^2$ is the average number of photons in a cavity.

\begin{figure}[h!]
	\center{\includegraphics[width=1\linewidth]{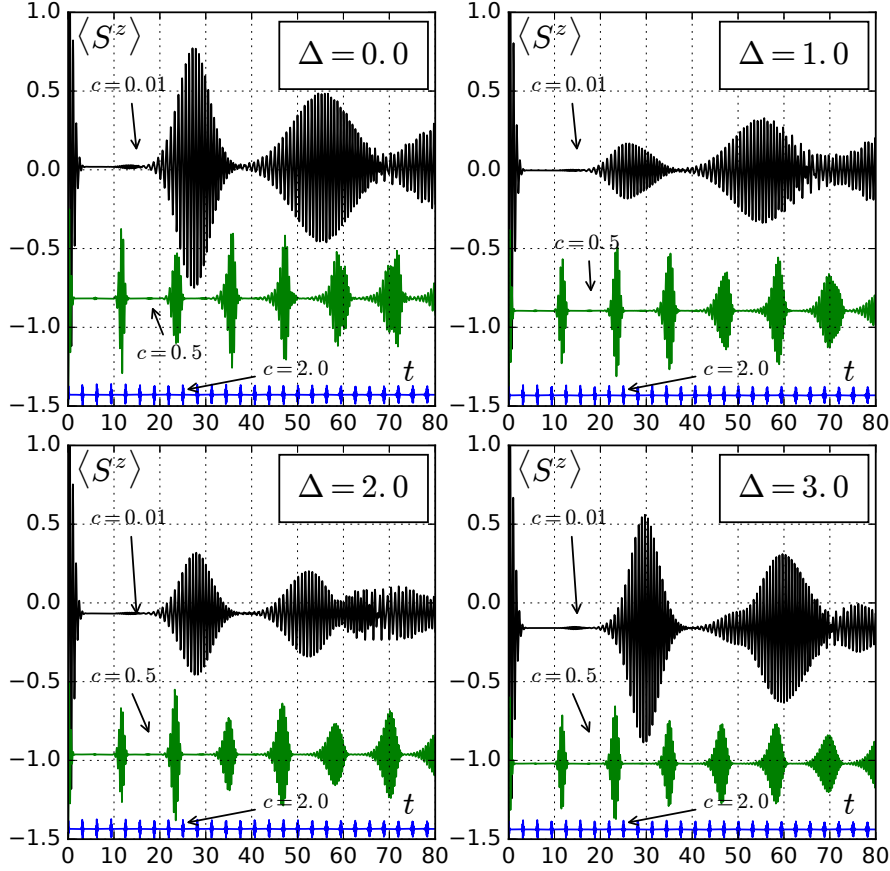}}
	\caption{Temporal evolution of the atomic inversion $\langle S^z\rangle$. The atoms are initially in the ground state with the total spin $S=\frac{3}{2}$, and the field is in coherent state with  $\langle n \rangle=20$. The parameters are equal to $c=0.01/0.5/2.0$ and $\Delta=0.0,1/0,2/0,3.0$.}
	\label{s15corr}
\end{figure}

The examples of numerical results for $\langle S^z\rangle(t)$ for different model parameters are given in the Fig.\ref{s15corr}. In sake of simplicity we consider the case when the cavity contains only 3 atoms, so the total spin $S$ changes in the range $[-\frac{3}{2},\frac{3}{2}]$. In the initial state $\mid {\bf\Phi}_0\rangle=\mid \alpha \rangle \mid S,-S\rangle$ all the atoms are in ground state. It's evident that the state $\mid {\bf\Phi}_0\rangle$ is not an eigenstate, so it rapidly decays in time. We see that the collapses  and revivals of Rabi oscillations for small coupling constants ($c=0.01$) are similar to those of the Jaynes-Cumming model \cite{fractional3} and time of the first revival is proportional to the average number of photons in a cavity  $t\sim <n>$.
As the effective coupling constant $c$ increases the revivals become more localized in time, their amplitudes and the time of the first revival decrease because of the dominating role of a Kerr medium.
The presence of the detuning affects the form of the revivals for small values of the effective coupling $c$.



Examples of neat evaluation of the $\langle S^z\rangle(t)$ are discussed in the next section.

\section{Numerical analysis of the Bethe equations and evaluation of the correlation functions}

\begin{figure}[h!]
	\center{\includegraphics[width=1\linewidth]{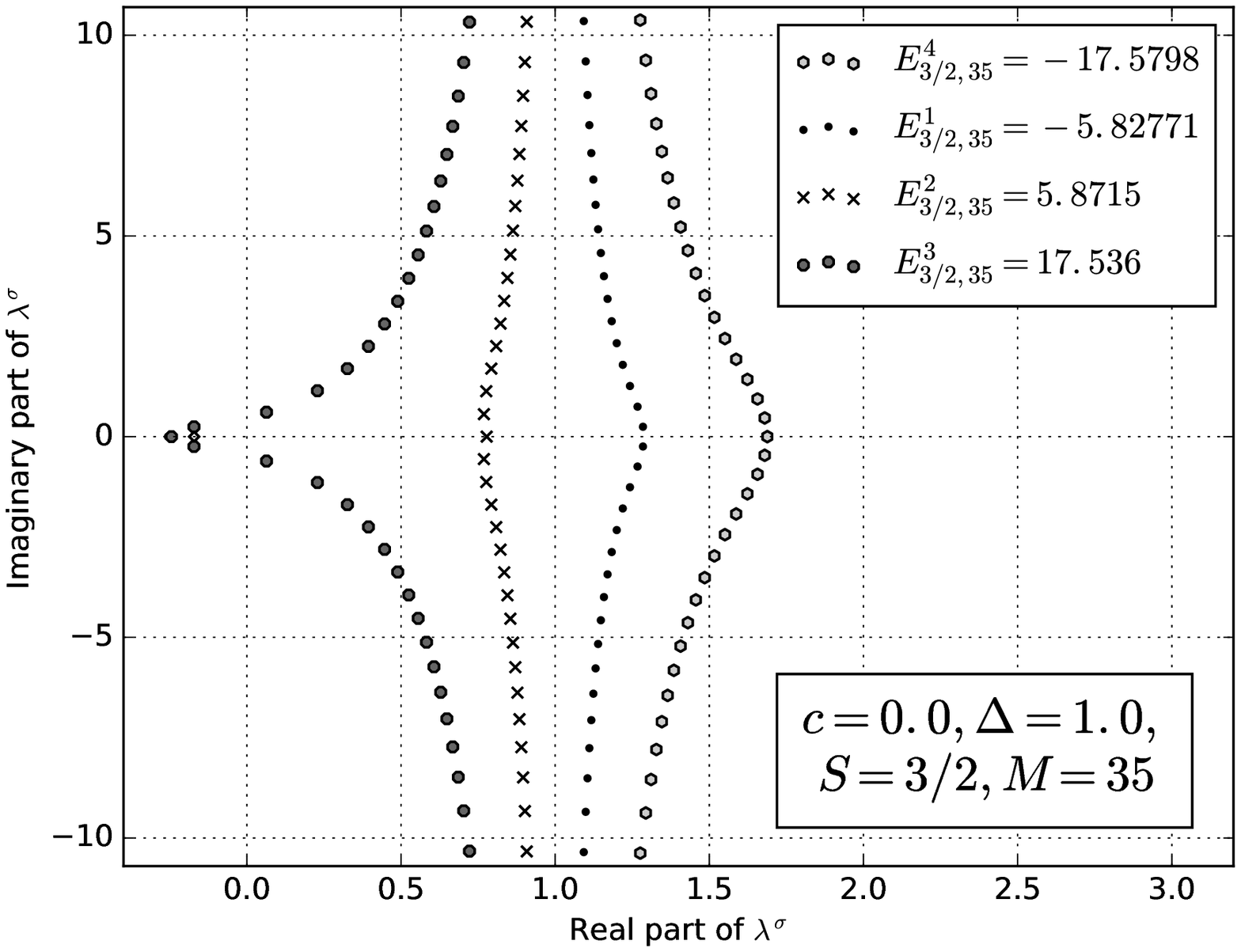}}
	\caption{The roots distribution $\lambda^\sigma_i$ of the Bethe equations (\ref{bethe}) with the parameters $c=0.0, M=35, \Delta=1.0$ and $S=\frac{3}{2}$.}
	\label{betheTC}
\end{figure}

The solutions of the Bethe equations (\ref{bethe}) give us a complete information about the system under consideration. 
The roots distribution of the Bethe equations (\ref{bethe}) and the spectrum for IGTC for the fixed values of parameters $c,\Delta,S,M$ are presented in Fig.(\ref{betheC05}), and in tables (\ref{tabBetheM4},\ref{tabBetheM15}) of the appendix. The roots distribution of the Bethe equations (\ref{bethetc}) and the spectrum for TC model for the fixed values of parameters $\Delta,S,M$ are presented in Fig.(\ref{betheTC}).

\begin{figure}[h!]
	\center{\includegraphics[width=1\linewidth]{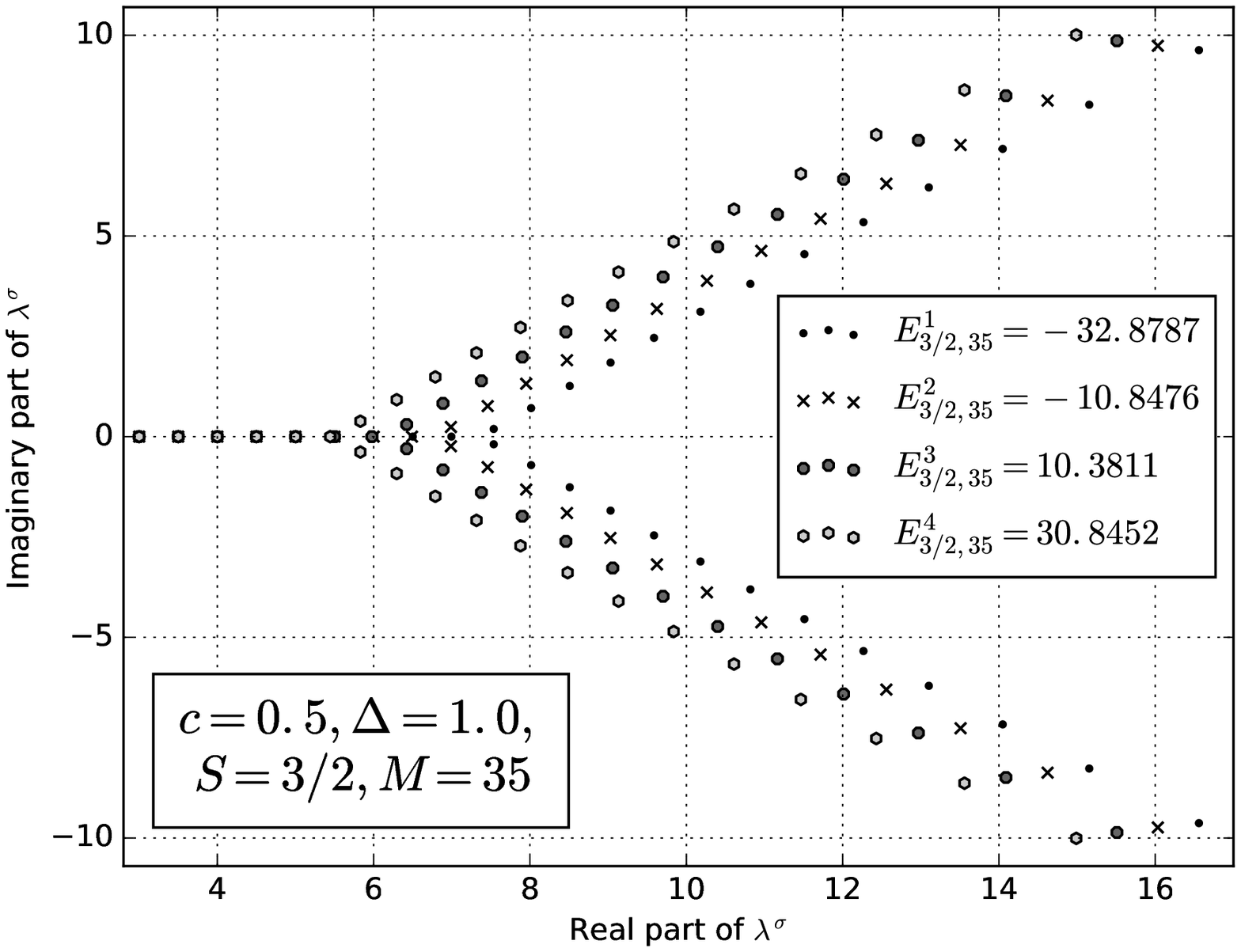}}
	\caption{The roots distribution $\lambda^\sigma_i$ of the Bethe equations (\ref{bethe}) with the parameters $c=0.5, M=35, \Delta=1.0$ and $S=\frac{3}{2}$. Note that not all the roots are presented in this figure.}
	\label{betheC05}
\end{figure}

\begin{figure}[h!]
	\center{\includegraphics[width=1\linewidth]{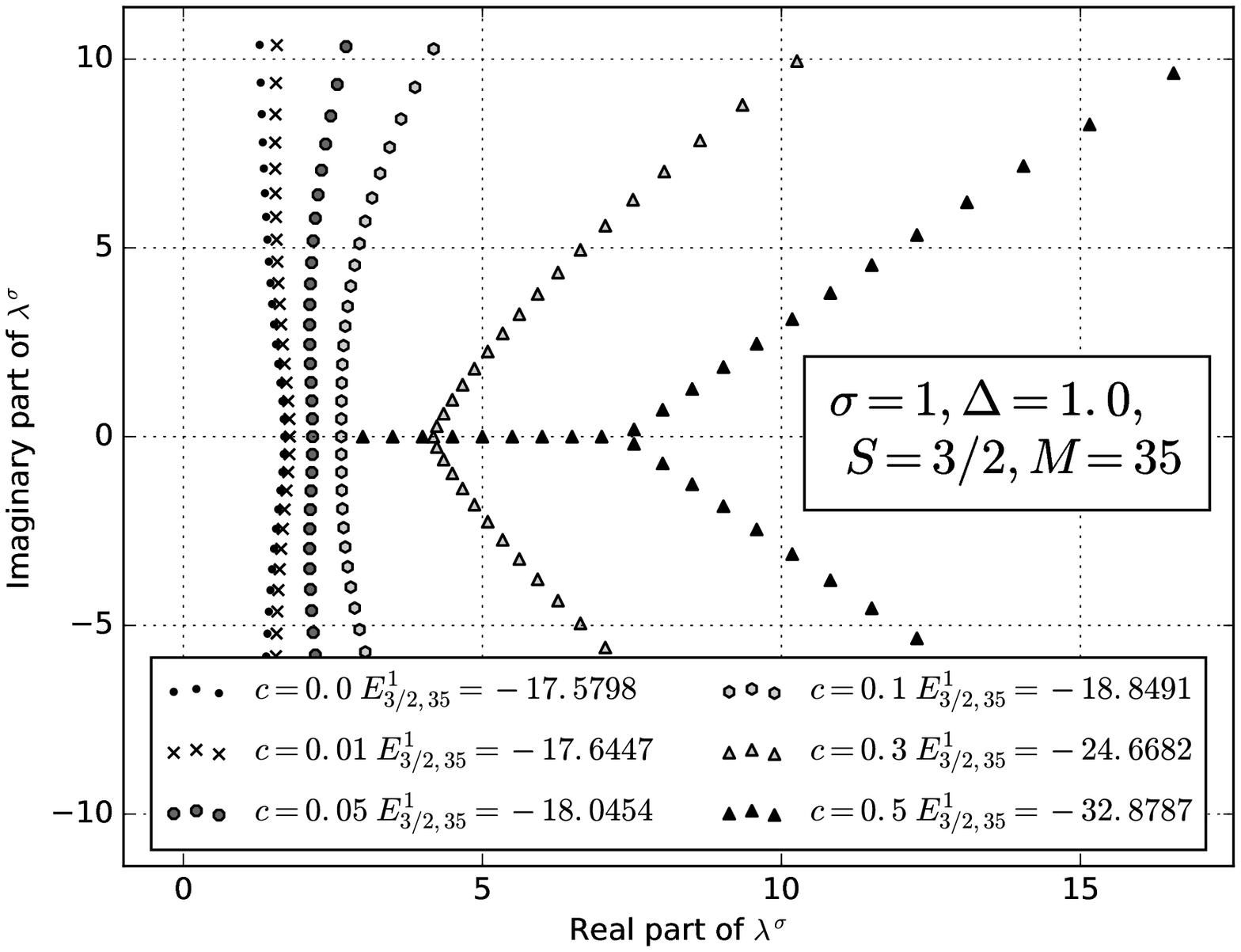}}
	\caption{Behaviour of the single solution $\{\bla^1\}$ of the Bethe equations (\ref{bethe}) for different $c$. Represented solutions refer to the ground state of the system with $\Delta=1.0, S=\frac{3}{2},M=35$}
	\label{evolGround}
\end{figure}

Solutions of the equations (\ref{bethe}) become close to the solutions of the equations (\ref{bethetc}), when $c\rightarrow 0$. The example of such transformation is presented in fig. \ref{evolGround}, for the solution $\{\bla^1\}$ which refers to the ground state of the system.

We use the iterative approach to the solution of Bethe equations (\ref{bethe}). Within this methods framework it is crucial to choose the correct initial assumption for the roots distribution in the first iteration.
The algorithm of finding the correct initial assumption is the following: ($I$) Solve the equations (\ref{bethetc}) for $\Delta=0.0$, taking the initial assumption in a string form with all roots having the same real part
 $\{a\pm ib\cdot 1,a\pm ib\cdot 2,a\pm ib\cdot 3,...\}$. By varying the parameters $a$ and $b$ we find the correct
  initial assumption for equations (\ref{bethetc}) which do not lead to non-physical solutions containing coinciding roots. ($II$) Use the solution of the equations (\ref{bethetc}) as an initial assumption for the equations (\ref{bethe}) with small $c$ and $\Delta$ (the optimal step can be found experimentally). ($III$) Vary $c$ and $\Delta$ slowly until it take the predetermined value.


In the Table (\ref{solutionsDifM}) we present the sets of solutions of the Bethe equations (\ref{bethe}) for this system, eigenenergies $E^\sigma_{3/2,M}$, and norms of the Bethe wavefunctions $\mathcal{N}^2_\sigma$. Using the data from the Table (\ref{solutionsDifM}) we evaluate the numerical values of the scalar products $f_{15}(\{\bla^\sigma\})$, $g_{15}(\{\bla^\sigma\})$, presented in the table (\ref{fmgNorm}), and we also evaluate the numericval values for the photonic transition elements: $A_{15,1}(\{{\bla}^{\sigma'} \},\{{\bla}^{\sigma} \})$ and $A^\dagger_{15,1}(\{{\bla}^{\sigma} \},\{{\bla}^{\sigma'} \})$, presented in Tables (\ref{ACoeffs}) and (\ref{ACoeffsConj}).

\section{Conclusions}

In this paper the QIM approach to the IGTC model was considered. We have obtained the determinant representation for the norm of the Bethe wavefunctions $\mathcal{N}^2_\sigma$ and for the transition elements of the photons $A_{M,n}(\{{\bla}^{\sigma'} \},\{{\bla}^{\sigma} \})$, $A^\dagger_{M,n}(\{{\bla}^{\sigma} \},\{{\bla}^{\sigma'} \})$. The knowledge of these elements allows to investigate any dynamical correlation function of the model provided the roots of Bethe equations are known. The  solutions of the Bethe equations for different values of parameters of the model were calculated numerically. The obtained results were applied for the evaluation of the atomic inversion $\langle\Phi(t)|S^z|\Phi(t)\rangle$.
%
\section{Acknowledgements}

The work was supported by the Russian Science Foundation (grant № 16-11-10218).

\section{Appendix}

Here we provide the solutions of the Bethe equations and the numerical values of the photonic transition elements and norms of wavefunctions.

\begin{table}[h!]
	\tiny
	\centering
	\caption{Solutions of the Bethe equations for $c=0.5, \Delta=1.0, S=\frac{3}{2},M=14,15$}
	\label{solutionsDifM}
	\begin{tabular}{cc}
		\begin{tabular}{|c|c|c|}
			\hline
			M & 14 & 15  \\
			\hline
			$\lambda^1_1$ & -0.239824 & -0.750193\\
			$\lambda^1_2$ & -0.750262 & -0.241867 \\
			$\lambda^1_3$ &  0.167747 &  0.176368  \\
			$\lambda^1_4$ &  1.773790 & 2.81293 $\pm$ 2.33351 I  \\
			$\lambda^1_5$ &  3.79826 $\pm$ 4.18216 I & 3.44429 $\pm$ 3.33488 I  \\
			$\lambda^1_6$ &  2.42263 $\pm$ 2.01687 I & 4.24773 $\pm$ 4.46706 I  \\
			$\lambda^1_7$ &  1.97772 $\pm$ 1.03823 I & 2.32525 $\pm$ 1.40143 I  \\
			$\lambda^1_8$ &  3.02276 $\pm$ 3.03848 I & 5.33330 $\pm$ 5.85846 I  \\
			$\lambda^1_9$ &  4.85600 $\pm$ 5.58194 I & 2.01400 $\pm$ 0.47729 I \\
			\hline
			$E^1_{3/2,M}$ & 14.3938 &15.2107 \\
			$\mathcal{N}^2_1$ & $1.32807\times 10^{22}$ & $1.11114\times 10^{25}$  \\
			\hline
			$\lambda^2_1$ & -0.73831 & -0.740738\\
			$\lambda^2_2$ & -0.36595 & -0.353435 \\
			$\lambda^2_3$ &  3.12892 + 1.92248 I &  2.59874  \\
			$\lambda^2_4$ &  2.73671 + 1.07169 I &  2.72731 + 0.62846 I  \\
			$\lambda^2_5$ &  2.51242 + 0.33328 I & 3.48714 + 2.22083 I  \\
			$\lambda^2_6$ &  3.67873 + 2.88682 I & 3.03737 + 1.37210 I  \\
			$\lambda^2_7$ &  4.41173 + 4.00060 I & 4.84837 + 4.28762 I  \\
			$\lambda^2_8$ &  5.43090 + 5.38255 I & 5.90050 + 5.66355 I  \\
			$\lambda^2_9$ &   & 4.07992 + 3.17973 I \\
			\hline
			$E^2_{3/2,M}$ & 4.80544 & 5.0842 \\
			$\mathcal{N}^2_2$ & $1.05391\times 10^{25}$ & $9.91835\times 10^{27}$  \\
			\hline
		\end{tabular} &
		\begin{tabular}{|c|c|c|}
			\hline
			M & 14 & 15  \\
			\hline
			$\lambda^3_1$ & 2.84973 & -0.823073\\
			$\lambda^3_2$ & -0.830452 & 2.98833 + 0.15896 I \\
			$\lambda^3_3$ &  5.01624 +3.90016 I & 4.10289 + 2.18817 I  \\
			$\lambda^3_4$ & 3.33298 +1.1325 I& 3.63948 +1.39331 I  \\
			$\lambda^3_5$ &  6.01243 +5.25218 I & 6.47453 +5.53095 I  \\
			$\lambda^3_6$ &  3.02186 +0.475188 I & 4.69008 +3.10476 I  \\
			$\lambda^3_7$ & 4.29896 +2.82574 I & 3.27731 +0.71051 I  \\
			$\lambda^3_8$ &  3.75156 +1.91465 I & 5.44343 +4.1804 I \\
			$\lambda^3_9$ & & \\
			\hline
			$E^3_{3/2,M}$ &-5.38734 & -5.65903 \\
			$\mathcal{N}^2_3$ & $4.49629\times 10^{27}$ & $4.9616\times 10^{30}$  \\
			\hline
			$\lambda^4_1$ &3.01453& 2.99698\\
			$\lambda^4_2$ & 3.25013 & 4.67624 +2.19107 I \\
			$\lambda^4_3$ &  3.87231 +1.18878 I&  3.44765 +0.216031 I  \\
			$\lambda^4_4$ &  5.60129 +3.85293 I & 4.19098 +1.42675 I \\
			$\lambda^4_5$ & 4.3222 +1.93711 I & 3.78925 +0.767612 I  \\
			$\lambda^4_6$ &  3.50608 +0.553672 I & 5.27176 +3.07648 I  \\
			$\lambda^4_7$ &  4.88324 +2.81231 I & 6.02349 +4.12297 I  \\
			$\lambda^4_8$ & 6.58851 +5.17388 I &7.04506 +5.44649 I  \\
			$\lambda^4_9$ &  & \\
			\hline
			$E^4_{3/2,M}$ & -16.3119 & -17.1358 \\
			$\mathcal{N}^2_4$ & $2.0875\times 10^{30}$ & $2.71893\times 10^{33}$  \\
			\hline
		\end{tabular}
	\end{tabular}
\end{table}

\begin{table}
	\centering
	\caption{Numerical values for the functions $f_{15}(\{\bla^\sigma\})$ and $g_{15}(\{\bla^\sigma\})$ for $c=0.5, \Delta=1.0, S=\frac{3}{2},M=15$.}
	\label{fmgNorm}
	\begin{tabular}{|c|c|c|}
		\hline
		$\sigma$ &  $f_{15}(\{\bla^\sigma\})$ & $g_{15}(\{\bla^\sigma\})$\\
		\hline
		1&  $-3.54225\times 10^9$ & $-5.71661\times 10^{11}$ \\
		2&  $1.88059\times 10^{12}$ & $1.31435\times 10^{13}$ \\
		3&  $-4.84651\times 10^{14}$ &$-1.47409\times 10^{14}$ \\
		4&  $8.58641\times 10^{16}$ &  $1.09485\times 10^{15}$\\
		\hline
	\end{tabular}
\end{table}

\begin{table}
	\centering
	\caption{Numerical values for the transition element $A_{15,1}(\{{\bla}^{\sigma} \},\{{\bla}^{\sigma'} \})$ for $c=0.5, \Delta=1.0, S=\frac{3}{2},M=15$.}
	\label{ACoeffs}
	\begin{tabular}{|c|c|c|c|c|}
		\hline
		$\sigma,\sigma'$ & 1 & 2 & 3 & 4\\
		\hline
		1& $-1.45983\times 10^{24}$ & $3.73937\times 10^{23}$ & $-6.08001\times 10^{22}$& $7.89756\times 10^{21}$ \\
		2& $1.65728\times 10^{24}$ & $-1.19478\times 10^{27}$ & $2.30675\times 10^{26}$& $-2.99671\times 10^{25}$\\
		3& $-2.81049\times 10^{23}$ & $1.28274\times 10^{27}$ &$-5.40508\times 10^{29}$& $8.05698\times 10^{28}$\\
		4& $3.75925\times 10^{22}$ & $-1.71635\times 10^{26}$ &  $5.62494\times 10^{29}$& $-2.6862\times 10^{32}$\\
		\hline
	\end{tabular}
\end{table}

\begin{table}
	\centering
	\caption{Numerical values for the transition element $A^\dagger_{15,1}(\{{\bla}^{\sigma} \},\{{\bla}^{\sigma'} \})$ for $c=0.5, \Delta=1.0, S=\frac{3}{2},M=15$.}
	\label{ACoeffsConj}
	\begin{tabular}{|c|c|c|c|c|}
		\hline
		$\sigma,\sigma'$ & 1 & 2 & 3 & 4\\
		\hline
		1& $-1.26067\times 10^{24}$ & $7.45656\times 10^{24}$ & $-2.78592\times 10^{25}$& $8.63194\times 10^{25}$ \\
		2& $6.8671\times 10^{22}$ & $-1.14315\times 10^{27}$ & $5.07159\times 10^{27}$& $-1.57159\times 10^{28}$\\
		3& $-5.57847\times 10^{20}$ & $5.87913\times 10^{25}$ &$-5.69246\times 10^{29}$& $2.02405\times 10^{30}$\\
		4& $3.41326\times 10^{18}$ & $-3.59844\times 10^{23}$ &  $2.70989\times 10^{28}$& $-3.0869\times 10^{32}$\\
		\hline
	\end{tabular}
\end{table}

\begin{table}
	\tiny
	\centering
	\caption{Example of the numerical solutions of the Bethe equations for $c=0.5,\Delta=0.1,M=4$, and different values of $S$. Note: this example is easy to reproduce.}
	\label{tabBetheM4}
	\begin{tabular}{|c|c|c|c|c|}
		\hline
		S & $\frac{1}{2}$ & 1 & $\frac{3}{2}$ & 2\\
		\hline
		$\lambda^1_1$ & 1.04808 $\pm$ 0.50837 i & 1.34952 $\pm$ 0.426617 i & 1.57716 $\pm$ 0.365265 i & 1.75806  $\pm$ 0.310326 i\\
		$\lambda^1_2$ & 1.29119 $\pm$ 1.89055 i & 1.53219 $\pm$ 1.61769 i & 1.75801  i $\pm$ 1.40880 i & 1.95484 $\pm$ 1.24021 i\\
		$\lambda^1_3$ & &  &  &\\
		\hline
		$E^1_{S,4}$ & -2.32855 & -4.46341 & -6.42034 & -8.22579 \\
		\hline
		$\lambda^2_1$ & -0.29968 $\pm$ 0.40156 i & 0.69235 $\pm$ 1.44650 i & 1.16794 $\pm$ 1.04222 i & 1.53492 $\pm$ 0.872592 i\\
		$\lambda^2_2$ & 0.435412 $\pm$ 2.03959 i & 0.87876 & 1.24467  & 1.52805\\
		$\lambda^2_3$ & & -0.73300 & -1.11421 & -1.55059\\
		\hline
		$E^2_{S,4}$ & 2.07855 & -0.23048 & -2.21633 & -3.84729\\
		\hline
		$\lambda^3_1$ & & -0.24857 $\pm$ 1.94860 i & -0.44967 $\pm$ 1.03001 i & 1.31263 $\pm$ 0.40260 i\\
		$\lambda^3_2$ & & -0.94836 $\pm$ 0.52991 i & -1.07433 & -1.34580 $\pm$ 0.584813 i\\
		$\lambda^3_3$ & &  & 0.987374  & \\
		\hline
		$E^3_{S,4}$ & & 3.69389 & 1.23630 & -0.733658\\
		\hline
		$\lambda^4_1$ & &  & -0.863032 $\pm$ 1.97470 i & -1.24456 $\pm$ 1.27796 i\\
		$\lambda^4_2$ & &  & -1.46216 $\pm$ 0.578535 i & 1.34077\\
		$\lambda^4_3$ & &  & & -1.66771\\
		\hline
		$E^4_{S,4}$ & & & 4.90038 & 2.01606\\
		\hline
		$\lambda^5_1$ & &  & & -1.91737 $\pm$ 0.611298 i\\
		$\lambda^5_2$ & &  & & -1.37797 $\pm$ 2.0319 i\\
		$\lambda^5_3$ & &  & & \\
		\hline
		$E^5_{S,4}$ & & &  & 5.79068 \\
		\hline
	\end{tabular}
\end{table}

\begin{table}[h]
	\tiny
	\centering
	\caption{Example of the numerical solutions of the Bethe equations for $c=0.5,\Delta=0.1,M=15$, and different values of $S$}
	\label{tabBetheM15}
	\begin{tabular}{|c|c|c|c|c|}
		\hline
		S & $\frac{1}{2}$ & 1 & $\frac{3}{2}$ & 2\\
		\hline
		$\lambda^1_1$ & 2.32741 $\pm$ 0.35373 i & 2.47583 $\pm$ 0.25997 i & 2.60227 $\pm$ 0.16887 i & 3.07694 $\pm$ 0.59944 i\\
		$\lambda^1_2$ & 2.65160 $\pm$ 0.95304 i & 2.80088 $\pm$ 0.82668 i & 2.94225 i $\pm$ 0.70956 i & 3.45953 $\pm$ 1.22965 i\\
		$\lambda^1_3$ & 3.05814 $\pm$ 1.65719 i & 3.19618 $\pm$  1.50389 i & 3.33006 $\pm$ 1.36210 i & 3.91847 $\pm$ 1.97124 i\\
		$\lambda^1_4$ & 3.56251 $\pm$ 2.46349 i & 3.68214 $\pm$ 2.28867 i & 3.80112 $\pm$ 2.12504 i & 4.48444 $\pm$ 2.84044 i\\
		$\lambda^1_5$ & 4.18708 $\pm$ 3.38509 i & 4.28484 $\pm$ 3.19415 i & 4.38431 $\pm$ 3.01280 i & 5.20525 $\pm$ 3.87615 i\\
		$\lambda^1_6$ & 4.97493 $\pm$ 4.46172 i & 5.04901 $\pm$ 4.25894 i & 5.12606 $\pm$ 4.06380 i & 6.19409 $\pm$ 5.19300 i\\
		$\lambda^1_7$ & 6.03978 $\pm$ 5.81056 i & 6.08796 $\pm$ 5.59900 i & 6.13958 $\pm$ 5.39312 i & 2.65482\\
		$\lambda^1_8$ & 2.06113 & 2.09262 & 2.09894 & 2.09986\\
		$\lambda^1_9$ & & & & 2.74737\\
		\hline
		$E^1_{S,15}$ & -5.46407 & -10.84627 & -16.15021 & -21.37955 \\
		\hline
		$\lambda^2_1$ & 1.77369 $\pm$ 0.29777 i & 1.53627 $\pm$ 0.81020 i & 2.12158 $\pm$ 0.12831 i & 2.57721 $\pm$ 0.55029 i\\
		$\lambda^2_2$ & 2.02909 $\pm$ 0.96717 i & 1.91551 $\pm$ 1.62851 i & 2.41032 $\pm$ 0.67232 i & 2.91733 $\pm$ 1.19437 i\\
		$\lambda^2_3$ & 2.42613 $\pm$ 1.72741 i & 2.42662 $\pm$  2.50233 i & 2.75830 $\pm$ 1.35172 i & 3.34863 $\pm$ 1.96357 i\\
		$\lambda^2_4$ & 2.94090 $\pm$ 2.57321 i & 3.06813 $\pm$ 3.46791 i & 3.21046 $\pm$ 2.14973 i & 3.90205 $\pm$ 2.86494 i\\
		$\lambda^2_5$ & 3.58330 $\pm$ 3.52257 i & 3.87683 $\pm$ 4.57537 i & 3.79058 $\pm$ 3.07152 i & 4.62248 $\pm$ 3.93212 i\\
		$\lambda^2_6$ & 4.39218 $\pm$ 4.61927 i & 4.96549 $\pm$  5.94766 i & 4.53977 $\pm$ 4.15231 i & 5.62050 $\pm$ 5.27839 i\\
		$\lambda^2_7$ & 5.48081 $\pm$ 5.98346 i & -0.49871 & 5.56837 $\pm$ 5.50753 i & 2.34232\\
		$\lambda^2_8$ & -0.26625 & -0.03260 & -0.81031 & 2.11376\\
		$\lambda^1_9$ & & 1.37064 & & -1.08752\\
		\hline
		$E^2_{S,15}$ & 5.21407 & -0.13668 & -5.38847 & -10.54503\\
		\hline
		$\lambda^3_1$ & & 1.97376 $\pm$ 0.21989 i & 1.81776 $\pm$ 0.620933 i & 2.05222 $\pm$ 0.48706 i\\
		$\lambda^3_2$ & & 2.22842 $\pm$ 0.80799 i & 2.11919 $\pm$ 1.37646 i & 2.32658 $\pm$ 1.16972 i\\
		$\lambda^3_3$ & & 2.59330 $\pm$ 1.52768 i & 2.57249 $\pm$ 2.23450 i & 2.73267 $\pm$ 1.99229 i\\
		$\lambda^3_4$ & & 3.07333 $\pm$ 2.35255 i & 3.17101 $\pm$ 3.19682 i & 3.28759 $\pm$ 2.94009 i\\
		$\lambda^3_5$ & & 3.68357 $\pm$ 3.29084 i & 3.94428 $\pm$ 4.30503 i & 4.02290 $\pm$ 4.04295 i\\
		$\lambda^3_6$ & & 4.46259 $\pm$ 4.38158 i & 5.00011 $\pm$ 5.67989 i & 5.04327 $\pm$ 5.41679 i\\
		$\lambda^3_7$ & & 5.52161 $\pm$ 5.74275 i & -0.32767 & 1.91606\\
		$\lambda^3_8$ & & -0.53645 & 1.70280  & -0.98428\\
		$\lambda^1_9$ & & & -0.74450 & -0.63740\\
		\hline
		$E^3_{S,15}$ & & 9.98296 & 4.71966 & -0.42488\\
		\hline
		$\lambda^4_1$ & &  & 0.92948 $\pm$ 0.62277 i & 1.44014 $\pm$ 0.33464 i\\
		$\lambda^4_2$ & &  & 1.36896 $\pm$ 1.54094 i & 1.61767 $\pm$ 1.18514 i\\
		$\lambda^4_3$ & &  & 1.89787 $\pm$ 2.44614 i & 2.04960 $\pm$ 2.10940 i\\
		$\lambda^4_4$ & &  & 2.54698 $\pm$ 3.42690 i & 2.64523 $\pm$ 3.10605 i\\
		$\lambda^4_5$ & &  & 3.35965 $\pm$ 4.54350 i & 3.41865 $\pm$ 4.23486 i\\
		$\lambda^4_6$ & &  & 4.45075 $\pm$ 5.92223 i & 4.47519 $\pm$ 5.62412 i\\
		$\lambda^4_7$ & &  & -0.75016 & -0.12358\\
		$\lambda^4_8$ & &  & 0.16566 & -1.00049\\
		$\lambda^1_9$ & & & -0.24192 & -0.480820\\
		\hline
		$E^4_{S,15}$ & & & 14.31902 & 9.11188\\
		\hline
		$\lambda^5_1$ & &  & & 0.20763 $\pm$ 0.68292 i\\
		$\lambda^5_2$ & &  & & 0.78426 $\pm$ 1.50981 i\\
		$\lambda^5_3$ & &  & & 1.35724 $\pm$ 2.41891 i\\
		$\lambda^5_4$ & &  & & 2.02265 $\pm$ 3.40599 i\\
		$\lambda^5_5$ & &  & & 2.84301 $\pm$ 4.52717 i\\
		$\lambda^5_6$ & &  & & 3.93844 $\pm$ 5.90929 i\\
		$\lambda^5_7$ & &  & & -0.30978\\
		$\lambda^5_8$ & &  & & -1.00065\\
		$\lambda^1_9$ &  & & & -0.43368\\
		\hline
		$E^5_{S,15}$ & & &  & 18.23759 \\
		\hline
	\end{tabular}
\end{table}


{\small\begin{thebibliography}{99}
		






\bibitem{Kulish1}
M. Chaichian, D. Ellinas, and P. Kulish,
\textit{Quantum algebra as the dynamical symmetry of the deformed Jaynes-Cummings model}, Phys. Rev. lett. \textbf{65}, 980 (1990).

\bibitem{rybin}
 A. Rybin, G. Kastelewicz, J. Timonen, N. Bogoliubov,
 \textit{The $su(1,1)$ Tavis-Cummings model},
 J. Phys. A: Math. Gen. \textbf{31}, 4705 (1998).

 \bibitem{babel}
 O. Babelon, D. Talalaev,
 \textit{On the Bethe Ansatz for the Jaynes-Cummings-Gaudin model},
  J.Stat.Mech.  \textbf{0706}, P06013 (2007).

\bibitem{braak}
D. Braak,
\textit{On the Integrability of the Rabi Model},
 Phys. Rev. Lett.  \textbf{107}, 100401 (2011).

 \bibitem{wvrk}
F.~ Wolf, F.~ Vallone, G.~ Romero, M.~ Kollar, E.~ Solano, D.~ Braak,
\textit{Dynamical correlation functions and the quantum Rabi model}, Phys. Rev. \textbf{A 87}, 023835 (2013).

 \bibitem{batch}
 M. T. Batchelor, Huan-Qiang Zhou,
 \textit{Integrability vs exact solvability in the quantum Rabi and Dicke models},
  Phys. Rev. A \textbf{91}, 053808 (2015).












	
		\bibitem{jc}
		E.~T.~Jaynes, F.~W.~Cummings,  \textit{ Comparison of quantum and semiclassical
			radiation theories with application to the beam maser}, Proc. IEEE \textbf{51}, 89 (1963).
		

		
		\bibitem{fractional1}
		F.~W.~Cummings,  \textit{Stimulated Emission of Radiation in a Single Mode}, Phys. Rev. A \textbf{140} 1051 (1965)
		
		\bibitem{fractional4}
		T. von Foerster, \textit{A comparison of quantum and semi-classical theories of the interaction between a two-level atom and the radiation field}, J. Phys. A \textbf{8}, 95 (1975).
		
		
		\bibitem{fractional3}
		J.~H.~Eberly, N.~B.~Narozhny, and J.~J.~Sanchez-Mondragon, \textit{Periodic Spontaneous Collapse and Revival in a Simple Quantum Model}, Phys. Rev. Lett. \textbf{44}, 1323 (1980).
		
		\bibitem{fractional5}
         N.~B.~Narozhny, J.~J.~Sanchez-Mondragon, and J.~H.~Eberly, \textit{Coherence versus incoherence: Collapse and revival in a simple quantum model}, Phys. Rev. A \textbf{23}, 236 (1981).		
		
\bibitem{scullyBook}
M. O. Scully, M. S. Zubairy, \textit{Quantum Optics}, Cambridge University Press, Cambridge, 1997.

\bibitem{haroch}
S. Haroche, D. Kleppner,
\textit{Cavity Quantum Electrodynamics},
Physics Today \textbf{42}, 24 (1989).		
		
		
		
		\bibitem{anniversary}
		A.~D.~Greentree, J.~Koch, J.~Larson, \textit{Fifty years of Jaynes-Cummings physics}, J. Phys. B \textbf{46}, 220201 (2013).
		
		\bibitem{circQED}
		S. M. Girvin, M. H. Devoret and R. J. Schoelkopf, \textit{Circuit QED and engineering charge-based superconducting qubits}, Phys. Scr. \textbf{T137}, 014012 (2009).
		
		\bibitem{JCrecent}
		Hudson Pimenta and Daniel F. V. James, \textit{Characteristic-function approach to the Jaynes-Cummings-model revivals}, Phys. rev. A. \textbf{94}, 053803 (2016).
		
		\bibitem{tc}
		M~Tavis, F.~W.~ Cummings, \textit{Exact solutions for an $N$-molecule-radiation-field Hamiltonian}, Phys. Rev. \textbf{170}, 279 (1968).
		
		
		\bibitem{lieb}
		K. Hepp, E. Lieb,
		\textit{On the Superradiant Phase Transition for Molecules in a Quantized Radiation Field: The Dicke Maser Model},
		Annals of Phys. (N.Y.) \textbf{76}, 360 (1973).
		
		\bibitem{jch}
		D.~G.~Angelakis, M.~F.~Santos and S.~Bose \textit{Photon-blockade-induced Mott transitions and XY spin models in coupled cavity arrays,} Phys. Rev. A \textbf{76}, 031805 (2007).
		
		\bibitem{rwarefuse}
		S.~Agarwal, S.~M.~Hashemi Rafsanjani, and J.~H.~Eberly \textit{Tavis-Cummings model beyond the rotating wave approximation: Quasidegenerate qubits}, Phys. Rev. A \textbf{85}, 043815 (2012).
		
		
\bibitem{rm}
B. M. Rodriguez-Lara,  H. M. Moya-Cessa,
\textit{The exact solution of generalized Dicke models via
Susskind-Glogower operators},
J. Phys. A: Math. Theor.  \textbf{46}, 095301 (2013).

\bibitem{btb}
H.~R.~Baghshahi, M.~K.~Tavassoly, A.~Behjat,
\textit{Entropy squeezing and atomic inversion in the k-photon
Jaynes-Cummings model in the presence of Stark shift
and Kerr medium: full nonlinear approach},  Chin. Phys. B \textbf{23},  074203 (2014).

\bibitem{hsk}
G. H. Hovsepyan, A. R. Shahinyan, G. Yu. Kryuchkyan
 \textit{Multiphoton blockades in pulsed regimes beyond the stationary limits},
  Phys. Rev. A \textbf{90}, 013839 (2014).

\bibitem{rs}
M. Rohith, C. Sudheesh
 \textit{Fractional revivals of superposed coherent states},
  J. Phys. B: At. Mol. Opt. Phys.  \textbf{47}, 045504 (2014).
		
\bibitem{kerr2}
		Imamoglu A., Schmidt H., Woods G., Deutsch M., \textit{Strongly Interacting Photons in a Nonlinear Cavity}, Phys. Rev. Lett. \textbf{79}, 8 (1997).		
		
			
		
		\bibitem{kerr3}
		Hai Wang, Goorskey D., Min Xiao, \textit{Controlling the cavity field with enhanced Kerr nonlinearity in three-level atoms}, Phys. Rev. A. \textbf{65}, 051802 (2002).
		
		\bibitem{kerr1}		
		Alexander K., Andre T., \textit{Strong reduction of laser power noise by means of a Kerr nonlinear cavity}, Phys. Rev. A \textbf{80}, 053801 (2009).
		
		\bibitem{wr}
		M.~J.~Werner, H.~Risken, \textit{Quasiprobability distributions for the cavitydamped
			Jaynes-Cummings model with an additional Kerr medium}, Phys. Rev. A \textbf{44}, 4623 (1991).

\bibitem{gj}
		P.~Gora, C.~Jedrzejek, \textit{Nonlinear Jaynes-Cummings model}, Phys. Rev. A \textbf{45},
		6816 (1992).
		
		\bibitem{jp}
		A.~Joshi, R.~R.~Puri, \textit{Dynamical evolution of the two-photon Jaynes-
			Cummings model in a Kerr-like medium}, Phys. Rev. A \textbf{45}, 5056 (1992).

			
\bibitem{nonlinJC}
P. Gora, C. Jedrzejec, \textit{Nonlinear Jaynec-Cummings model}, Phys. Rev. A. \textbf{45}, 6816 (1992).
\bibitem{joshi}
		Amitabh Joshi and R. R. Puri, \textit{Dynamical evolution of the two-photon Jaynes-Cummings model in a Kerr-like medium}, Phys. rev. A \textbf{45}, 5056 (1992).
		
\bibitem{nonlinRes1}T. V. Gevorgyan, A. R. Shahinyan, G. Yu. Kryuchkyan, \textit{Quantum interference and sub-Poissonian statistics for time-modulated driven dissipative
nonlinear oscillators}, Phys. rev. A \textbf{79}, 053828 (2009).

\bibitem{nonlinRes2}G. H. Hovsepyan, A. R. Shahinyan, \textit{Phase locking and quantum statistics in a parametrically driven nonlinear resonator}, Phys. rev. A \textbf{93}, 043856 (2016).
		
		
\bibitem{milburn}G. J. Milburn and J. Corney, \textit{Quantum dynamics of an atomic Bose-Einstein condensate in a double-well potential}, Phys. rev. A \textbf{55}, 4318 (1997).

\bibitem{boseHub}M. K. Olsen and C. V. Chianca, \textit{Quantum correlations in pumped and damped Bose-Hubbard dimers}, Phys. rev. A. \textbf{94}, 043604 (2016).

\bibitem{nmb}
N.M. Bogoliubov, \textit{Time-dependent correlation functions of the two-mode Bose-Hubbard model},
J. Math. Sci.  \textbf{213} , 662 (2016).

		
				
		\bibitem{bkul}
     N. M. Bogoliubov, P. P. Kulish
      \textit{Exactly solvable models of quantum nonlinear optics},
      J. Math. Sci. (New York) \textbf{192}, 14 (2013).	

		\bibitem{gaud}
		M. Gaudin, \textit{La fonction d'onde de Bethe}, Masson, Paris, 1983.
		
		\bibitem{bbt}
		N.~M.~Bogoliubov, R.~K.~Bullough, J.~Timonen,
		\textit{Exact solution of generalised Tavis-Cummings models in quantum optics},
		J. Phys. A \textbf{29}, 6305 (1996).		
		
		\bibitem{fad}
		L.~D.~Faddeev,
		\textit{Quantum completely integrable models of field theory}, Sov. Sci. Rev. Math. C \textbf{1}, 107 (1980).
		
		\bibitem{Kulish2}
		P. P. Kulish, E. K. Sklyanin, \textit{Quantum spectral transform method recent developments}, Lect Notes Phys. \textbf{151}, 61 (1982).		
		
		\bibitem{kbi}
		V.~E.~Korepin, N.~M.~Bogoliubov, A.~G.~Izergin,
		\textit{Quantum Inverse Scattering Method and Correlation
			Functions}, Cambridge University Press, Cambridge, 1993.
		
		

		
		\bibitem{korepinnorm}
		V.~E.~Korepin, \textit{Calculation of Norms of Bethe Wave Functions}, Commun. Math. Phys. \textbf{86}, 391 (1982).
		
		\bibitem{odeim}
		P.~Dorey, C.~Dunning and R.~Tateo \textit{The ODE/IM correspondence}, J. Phys. A \textbf{40}, 32 (2007).
		
		
	
		
		\bibitem{gaudinmodel}
		A.~Faribault, O.~El.~Araby, C.~Strater, and V.~Grivtsev \textit{Gaudin models solver based on the Bethe ansatz/ordinary differential equations correspondence}, Phys. Rev. B \textbf{83}, 235124 (2011).
		
		
		\bibitem{strings}
		R.~Hagemans and Jean-Sebastien Caux \textit{Deformed strings in the Heisenberg model}, J. Phys. A \textbf{40}, 14605 (2007).
		

		
		\bibitem{vladimirov} A.~A.~Vladimirov, \textit{Proof of the invariance of the Bethe-ansatz solutions under complex conjugation}, Theoretical and Mathematical Physics \textbf{66}, 102 (1986).
		
		\bibitem{macd}
		I.~G.~Macdonald,
		{\it Symmetric Functions and Hall Polynomials}, Oxford University Press, Oxford, 1995.
		
		
		
		\bibitem{sl}
		N.~A.~Slavnov, \textit{Calculation of scalar products of wave functions and form factors in the framework of the alcebraic Bethe ansatz}, Theor. Math. Phys \textbf{79}, 502 (1989).
		
		\bibitem{kit}
		N.~Kitanine, J.~M.~Maillet, V.~Terras,
		\textit{Form factors of the XXZ Heisenberg spin-1/2 finite chain},
		Nucl. Phys. B \textbf{516}, 647 (1999).
		
		\bibitem{kit2}
		 N. Kitanine, K. Kozlowski, J. M. Maillet, N. A. Slavnov and V. Terras, \textit{On correlation functions of integrable models associated with the six-vertex R-matrix}, J. Stat. Mech., 01022 (2007).

		
\end{thebibliography}}
\end{document}